\documentclass[runningheads]{llncs}

\usepackage[T1]{fontenc}
\usepackage{graphicx}
\usepackage{siunitx}
\usepackage{amsmath,amssymb}
\usepackage{booktabs}
\usepackage{multirow}
\usepackage{xspace}
\usepackage{xcolor}
\usepackage{siunitx}
\usepackage{subcaption}
\usepackage{algorithm}
\usepackage[noend]{algpseudocode}
\usepackage{tikz}
\usetikzlibrary{arrows.meta, positioning, calc, fit, backgrounds}
\usepackage{listings}
\usepackage{url}
\usepackage[nospace]{cite}
\usepackage[hidelinks]{hyperref}

\usepackage{orcidlink}
\renewcommand{\orcidID}[1]{\unskip\,\orcidlink{#1}}

\usepackage{color}

\graphicspath{{./graphics/}}

\newcommand{\gpap}{\textsc{Gpap}\xspace}
\newcommand{\gpaps}{\mbox{\textsc{Gpap-S}}\xspace}

\newcommand{\sumword}{sum-packed word\xspace}

\newcommand{\noblecount}{\textsc{NobleCount}\xspace}
\newcommand{\wave}{\textsc{Wave}\xspace}

\newcommand{\cmsgen}{\textsc{CMSGen}\xspace}
\newcommand{\demotic}{\textsc{Demotic}\xspace}

\newcommand{\recurrentfull}{\textsc{Recurrent-CircuitSAT}\xspace}
\newcommand{\recurrent}{\textsc{Recurrent}\xspace}
\newcommand{\diffsampler}{\textsc{DiffSampler}\xspace}

\newcommand{\spur}{\textsc{Spur}\xspace}
\newcommand{\barbarik}{\textsc{Barbarik}\xspace}

\newcommand{\abctool}{\textsc{Abc}\xspace}
\newcommand{\triton}{\textsc{Triton}\xspace}

\newcommand{\turbosat}{\textsc{TurboSAT}\xspace}
\newcommand{\fastfouriersat}{\textsc{FastFourierSAT}\xspace}
\newcommand{\nanosat}{\textsc{NanoSAT}\xspace}
\newcommand{\turbo}{\textsc{Turbo}\xspace}
\newcommand{\runin}[1]{\textit{#1}}
\newcommand{\bcp}{BCP\xspace}
\newcommand{\orig}{original\xspace}
\newcommand{\cnf}{CNF\xspace}
\newcommand{\satcount}{\ensuremath{\mathit{cnt}}}
\newcommand{\litsum}{\ensuremath{\mathit{sum}}}
\newcommand{\occ}{\ensuremath{\mathit{occ}}}
\newcommand{\pkt}{\ensuremath{\mathit{st}}}
\newcommand{\shift}{\ensuremath{r}}
\newcommand{\litset}{\ensuremath{L}}
\newcommand{\support}{\ensuremath{S}}

\newcommand{\nrows}{\ensuremath{K}}
\newcommand{\nclauses}{\ensuremath{M}}
\newcommand{\nvars}{\ensuremath{N}}
\newcommand{\ndec}{\ensuremath{B}}
\newcommand{\depth}{\ensuremath{T}}
\newcommand{\wlanes}{\ensuremath{W}}

\newcommand{\implps}{implications/s\xspace}
\newcommand{\e}[1]{\ensuremath{\times 10^{#1}}}

\definecolor{darkgreen}{rgb}{0.0, 0.35, 0.0}

\newcommand{\ys}[1]{}
\newcommand{\todo}[1]{}

\begin{document}

\title{SAT Sampling at Scale via GPU-based Boolean Constraint Propagation}
\titlerunning{SAT Sampling at Scale via GPU-based BCP}

\author{Alexander Nadel\orcidID{0000-0003-4679-892X} \and
        Yogev Shalmon\orcidID{0009-0004-3720-4004} \and
        Ofer Strichman\orcidID{0000-0001-9169-3751}}
\authorrunning{A. Nadel et al.}
\institute{Faculty of Data and Decision Sciences, Technion, Haifa, Israel\\
\email{\{alexandernad,ofers\}@technion.ac.il}, \email{yogevshalmon@campus.technion.ac.il}}

\maketitle

\begin{abstract} 

Constrained-random verification tests hardware designs by sampling solutions of Boolean formulas. Existing samplers largely follow one of two approaches: CPU-based SAT samplers (e.g., \cmsgen{}), which remain effective even at rare targets but offer limited throughput, and GPU-based gradient samplers, which achieve high throughput at high-probability targets but collapse at rare ones. 
We introduce \gpaps{}, the first GPU sampler built on Boolean constraint propagation (BCP). At its core is \gpap{}, a new GPU engine for BCP that efficiently propagates many independent partial assignments concurrently on a shared Boolean formula. This capability allows \gpaps{} to combine the rare-target robustness of SAT-based sampling with the throughput of GPU parallelism by exploring thousands of decide-and-propagate trajectories simultaneously. 
On the benchmark families used to evaluate prior GPU gradient samplers, \gpaps{} is $2.2$--$18.7\times$ faster at the high-probability targets considered in those evaluations. At rare targets, where those samplers often fail to generate any samples, \gpaps{} achieves a geometric-mean speedup of roughly two orders of magnitude. Across all benchmarks, \gpaps{} is also $83.4$--$1{,}571\times$ faster than the CPU-based \cmsgen{}. 
This throughput does not come at the expense of sample quality: under \barbarik{}, the statistical tester used to evaluate \cmsgen{}, \gpaps{} is accepted on every instance we test, as is \cmsgen{}. 
We release \gpap{} as a reusable GPU library for high-throughput BCP.

\end{abstract}


\section{Introduction}
\label{sec:intro}
Constrained-random verification is one of the backbones of hardware
verification~\cite{naveh_constraint_2006,kitchen_stimulus_2007}: before a design is manufactured,
it is simulated on many random input stimuli, drawn under constraints that keep them legal and
steer them toward the scenarios worth testing, including corner cases that random inputs rarely
reach. The constraints can be encoded as a Boolean formula, whose solutions are the legal stimuli~\cite{chakraborty_unigen_2014}. When
they constrain the design's outputs, as in our evaluation, the formula also encodes the circuit,
and a solution is a stimulus that drives it to the required
outputs~\cite{ardakani_demotic_2025,ardakani_recurrent_2025}. The task is thus
to sample solutions: many and fast, since coverage grows with the number of stimuli; spread out
rather than clustered; and each one valid, since a stimulus that violates the constraints is not
a legal input at all. The same
need for many random solutions of a constraint set arises in software
testing~\cite{dutra_quicksampler_2018}.

Today's samplers largely follow one of two approaches, and they prioritize different sides of
this task: SAT-based samplers, which run on a central processing unit (CPU) and put deductive
power first, and gradient-based samplers, which run on a graphics
processing unit (GPU) and put throughput first.

The SAT-based samplers build on conflict-driven clause learning (CDCL) SAT
solvers~\cite{marques-silva_grasp_1999,moskewicz_chaff_2001}, which today handle
industrial formulas with millions of clauses in practice~\cite{fichte_silent_2023,katebi_empirical_2011}.
The core CDCL algorithm repeats
one loop: \emph{decide} a value for an unassigned variable, \emph{propagate} its consequences by
Boolean constraint propagation (\bcp), which repeatedly forces literals implied by the unit-clause
rule until fixpoint or a conflict, and, on a conflict, \emph{analyze} it into a learned
clause and \emph{backtrack}. A SAT-based sampler obtains each sample from one such solver run.
\cmsgen{}~\cite{golia_cmsgen_2021}, for instance, is a CDCL solver with randomized heuristics: it
picks the next variable and its polarity uniformly at random, propagates, learns from conflicts as
usual, and returns the model it reaches. \cmsgen{} targets not only validity but also a distribution close enough to uniform to provide meaningful verification coverage, which it terms \emph{uniform-like}.

A CDCL solver spends most of its runtime in \bcp{}~\cite{moskewicz_chaff_2001,zhang_cache_2004}; samplers built on CDCL therefore inherit this bottleneck. Decades of research have refined \bcp{} for CPUs, whose small number of sophisticated cores are designed to execute a single sequential program as quickly as possible, yet \bcp{} remains the dominant cost. An alternative is to change the hardware rather than the algorithm. GPUs make the opposite design trade-off: thousands of simpler cores that deliver high throughput when many independent tasks are available to hide memory latency. A single \bcp{} run, however, offers little obvious task parallelism. Several efforts have nevertheless attempted to accelerate \bcp{} on GPUs. As we discuss in Section~\ref{sec:related}, these works report speedups only relative to their own simplified propagators rather than the propagation schemes used in state-of-the-art CPU solvers, and none targets sampling.

A different line of work keeps the GPU but abandons propagation: several recent tools relax the formula into a continuous function and search it with gradient-based methods, rounding the result to Boolean assignments.
This yields high throughput, but a rounded assignment counts only if it satisfies the formula. At rare targets, where random inputs seldom reach the desired region, few rounded assignments are valid, and throughput drops by orders of magnitude. Among these samplers, \diffsampler{}~\cite{ardakani_diffsampler_2024} and its high-throughput successor~\cite{ardakani_hts_2025} target formulas in conjunctive normal form (\cnf{}), whereas \demotic{}~\cite{ardakani_demotic_2025} and \recurrentfull{}~\cite{ardakani_recurrent_2025} (\recurrent{} for short) target circuits. We compare against the latter two on circuits from their benchmark families.

We present \gpaps{}, the first GPU sampler built on SAT propagation, combining the strengths of both existing approaches: the rare-target robustness of SAT-based samplers and the throughput of GPU parallelism. Like SAT-based samplers, \gpaps{} advances through \bcp{}, so every trajectory that completes yields a model; unlike them, it executes thousands of trajectories concurrently on a GPU. \gpaps{} is built on \gpap{} (GPU-Parallel (partial) Assignment Propagation), a novel parallel \bcp{} engine. \gpap{} propagates $\nrows$ independent partial assignments to fixpoint concurrently on a shared formula. Each assignment follows its own decide-and-propagate trajectory, while conflict analysis and backtracking are eliminated entirely: a trajectory that reaches a conflict is simply discarded. To enable this design, we use a per-clause data structure that combines a counter with a running sum of the surviving literals' indices. This structure is a variant of a pairing that appeared in early SAT solvers and local-search methods~\cite{biere_evolution_2004,balint_improving_2014}, that was largely overlooked thereafter, and that we recently revived in \noblecount{}~\cite{noblecount_fmcad_2026}. As we show, it is particularly well suited to GPU execution.

We focus our evaluation on combinational and sequential circuits.
Although \gpaps{} itself operates on plain \cnf{}, constrained-random verification is our motivating application, and circuits are its natural domain. In this setting, a sample is an input stimulus, a \emph{target} is an output pattern the circuit is constrained to reach, and a \emph{rare} target is one that random stimuli seldom hit, making satisfying stimuli scarce. We compare against \cmsgen{} and the GPU gradient samplers on 41 circuits drawn from the benchmark families used in the latter's evaluations. For each circuit, we evaluate both the target used in the original papers and a rare target of our own construction, running all tools on the same machine.

Across all benchmarks, \gpaps{} samples $83.4$--$1{,}571\times$ faster than \cmsgen{} in geometric mean, depending on the benchmark set, target, and unrolling depth. Relative to the GPU gradient samplers, performance depends primarily on target rarity. At the targets used in their original evaluations, \gpaps{} is $2.2\times$ faster than \demotic{} and $8.0$--$18.7\times$ faster than \recurrent{}. At rare targets, \gpaps{} is roughly two orders of magnitude faster than \recurrent{}, while \demotic{} fails to produce samples on most instances (Section~\ref{sec:eval}). This throughput does not come at the expense of sample quality. Under \barbarik{}~\cite{chakraborty_barbarik_2019}, the statistical tester used to evaluate \cmsgen{}, \gpaps{} is accepted on every circuit instance we consider, as is \cmsgen{}.

\paragraph{Our Contributions.} 

\begin{itemize} 

\item \textbf{\gpaps{}, an open-source GPU sampler built on propagation.}\footnote{\url{https://github.com/yogevshalmon/gpap}} Every trajectory that completes yields a model by construction, with no rounding or repair step required (Section~\ref{sec:sampling}). 

\item \textbf{\gpap{}, the parallel \bcp{} engine behind it.} \gpap{} propagates up to the order of $10^{5}$ partial assignments in a single GPU launch, all against a shared read-only formula, and executes each decide-and-propagate trajectory entirely within one kernel launch (Section~\ref{sec:engine}). 

\item \textbf{A lock-free propagation primitive.} Each clause stores its state in a single 32-bit word, allowing literal falsification with one atomic instruction. The returned value immediately indicates whether the clause became unit, and on which literal, or reached a conflict (Section~\ref{sec:engine:word}). 

\item \textbf{A general-purpose \bcp{} library.} We release \gpap{} as an open-source library, in the same repository, with C++ and Python interfaces. The same batched-propagation primitive applies wherever a solver evaluates many candidate assignments through propagation, including probing, SAT sweeping, and look-ahead. We hope it serves as a building block for future GPU SAT research beyond this work (Section~\ref{sec:engine:api}). 

\end{itemize}

The remainder of the paper is organized as follows. Section~\ref{sec:prelim} introduces the necessary preliminaries, and Section~\ref{sec:related} surveys related work. Section~\ref{sec:engine} presents \gpap{}, and Section~\ref{sec:sampling} the sampler built upon it. Section~\ref{sec:eval} describes the experimental methodology and reports the results, and Section~\ref{sec:conclusion} concludes.

\section{Preliminaries}
\label{sec:prelim}

\subsection{SAT, Propagation, and Projected Sampling}

A \emph{CNF formula} $F = C_1 \wedge \dots \wedge C_{\nclauses}$ over a set of Boolean variables
$V = \{x_1,\dots,x_{\nvars}\}$ is a conjunction of $\nclauses$ \emph{clauses}, each clause being a
disjunction of \emph{literals}, where a literal is a variable $x$ or its negation $\neg x$. A
\emph{partial assignment} $\alpha$ maps a subset of $V$ to $\{0,1\}$. A literal is \emph{satisfied}
if $\alpha$ maps it to 1, \emph{falsified} if $\alpha$ maps it to 0, and \emph{free} otherwise.
Under $\alpha$, a clause is \emph{satisfied} if one of its literals is; \emph{conflicting} if all of
its literals are falsified; and \emph{unit} if exactly one literal is free and all others are
falsified, in which case that literal is \emph{forced}: every extension of $\alpha$ that
satisfies $F$ must satisfy it.

\emph{Boolean constraint propagation} (\bcp) is the closure operation that repeatedly asserts forced
literals until no clause is unit, or until some clause becomes conflicting; each such assertion is
an \emph{implication}. We rely throughout on the confluence of \bcp{} in the following sense: for a given $\alpha$, the order in which unit clauses are processed affects neither the resulting fixpoint nor whether a conflict is reached, although it may affect which clause witnesses the conflict. We represent a partial assignment $\alpha$ by the set $\litset$ of literals it satisfies. Applying \bcp{} to $\litset$ in $F$ either derives a conflict, proving that $\alpha$ cannot be extended to a model of $F$, or reaches a fixpoint that extends $\litset$ with all the implications it derives.

\paragraph{Literal Indexing.}
We identify literals with integers in $[0, 2\nvars)$: the positive literal of $x_v$ is $2(v-1)$ and
its negation is $2(v-1)+1$, so negation is a single bit flip, $\neg\ell = \ell \oplus 1$, and the two polarities
of a variable are adjacent. The \emph{negated occurrence list} $\occ[\ell]$ is the list of clauses
in which $\neg \ell$ occurs, and therefore exactly the clauses in which asserting $\ell$
falsifies a literal.

\paragraph{Per-Clause Bookkeeping.} Whenever a literal in a clause is falsified, propagation must answer two questions: has the clause become conflicting, and has it become unit, in which case the forced literal must also be identified. CDCL solvers typically address these questions with the two-watched-literals scheme (2WL)~\cite{moskewicz_chaff_2001}. Each clause is watched by two non-falsified literals, so the clause is revisited only when one of them is falsified. The solver then searches the clause for a non-falsified replacement and moves the clause to that literal's watch list. The alternative we build on uses no watches. Instead, each clause stores two values: a \emph{counter} and a \emph{key}. Propagation reads and updates only these values, never the clause's literals. The scheme dates back to early SAT solvers and local-search methods~\cite{biere_evolution_2004,balint_improving_2014}. The counter $\satcount$ records the number of literals not yet falsified, which we call the clause's \emph{surviving} literals. Thus, $\satcount = 0$ indicates a conflict, while $\satcount = 1$ indicates that the clause has become unit. The key stores the XOR of the indices of all surviving literals. Falsifying a literal removes its index with a single XOR operation, and when only one literal survives, the key itself equals that literal's index, directly identifying the forced literal. This simplicity comes with two drawbacks relative to 2WL. First, assigning a literal $\ell$ updates every clause in $\occ[\ell]$, whereas 2WL visits only clauses in which the falsified literal is currently watched. Second, solvers that backtrack must undo these updates, whereas watched literals require no undoing.

\begin{example}
Take the clause $(x_2 \vee \neg x_5 \vee x_7)$. Its literals have indices $2(2-1) = 2$,
$2(5-1)+1 = 9$, and $2(7-1) = 12$, so it starts with $\satcount = 3$ and key
$2 \oplus 9 \oplus 12 = 7$. Assigning $x_5=1$ falsifies $\neg x_5$, leaving $\satcount = 2$ and key
$7 \oplus 9 = 14$. Assigning $x_2=0$ then leaves $\satcount = 1$ and key $14 \oplus 2 = 12$:
the clause is unit, and $12$ is the index of $x_7$, the forced literal.
\end{example}

\paragraph{Projected Solution Sampling.} Let $\support \subseteq V$ be a designated \emph{support}. The projection of a model onto $\support$ is its restriction to those variables. A \emph{projected sampler} draws from the set of all such projections. It is \emph{uniform} if every projection is equally likely. Section~\ref{sec:related:sampling} discusses the cost of guaranteeing such uniformity.

\paragraph{Circuits.} Our benchmarks are circuits rather than plain CNF. A sequential circuit is unrolled from its reset state for $\depth$ cycles (its \emph{depth}), constrained so that selected final-cycle outputs match a chosen pattern, and encoded into CNF~\cite{tseitin_complexity_1983}; a combinational circuit is the special case $\depth = 1$. The support consists of all primary inputs across all cycles, so a sample is an input stimulus that drives the circuit to the chosen output pattern, which we call the \emph{target}. A \emph{random} target is produced by a random stimulus and is typically easy to reach. A \emph{rare} target is the least frequent pattern observed in a large simulation batch; by construction it is reachable and therefore satisfiable. This is also the setting considered by recent GPU circuit samplers~\cite{ardakani_demotic_2025,ardakani_recurrent_2025}.

\subsection{The GPU Execution Model} \label{sec:prelim:gpu} A GPU executes a \emph{kernel}: a single program launched across many threads, which we call \emph{lanes}, following GPU vendor terminology. The device comprises many \emph{streaming multiprocessors (SMs)}; a kernel distributes its lanes across them, and each SM schedules lanes in fixed groups of 32, called \emph{warps}, that share one instruction stream and execute in lockstep. An SM keeps several warps \emph{resident} simultaneously and hides the hundreds of cycles of memory latency by switching to another warp while one waits. Throughput therefore depends on \emph{occupancy}: the number of warps, and hence of our \emph{jobs} (one per partial assignment, Section~\ref{sec:engine:structure}), resident at once. Four properties of this model shape our design. 

\begin{description} 

\item[Saturation.] Too few jobs leave the machine underutilized, regardless of arithmetic speed. Maintaining efficiency therefore requires enough independent jobs to keep the GPU busy. 

\item[Compactness.] A job's private state occupies memory for its entire lifetime. Keeping this state compact, especially the per-clause bookkeeping, increases the number of jobs that fit concurrently; thus \emph{bytes per clause} directly determine how many jobs fit. 

\item[Divergence.] Divergence within a warp is serialized, reducing throughput. Since each job occupies its own warp, only lanes within the same job can diverge; different jobs do not interfere. 

\item[Atomics over locks.] Atomic read-modify-write instructions execute in hardware and return the previous value. Concurrent accesses are serialized by the memory system, eliminating the need for locks. Algorithms that base decisions on the returned values require no further synchronization. 

\end{description}

\section{Related Work}
\label{sec:related}

\subsection{Solution Sampling}
\label{sec:related:sampling}

Guaranteeing uniformity is expensive: for SAT, almost-uniform generation and approximate model
counting are polynomially inter-reducible~\cite{jerrum_random_1986}, and samplers with such
guarantees either make many SAT calls on the formula conjoined with random XOR
constraints~\cite{chakraborty_unigen_2014,soos_unigen3_2020}, or must first compile the formula
to \mbox{d-DNNF}~\cite{sharma_kus_2018} or count its models exactly~\cite{achlioptas_spur_2018}. A
practical line of work targets samplers that are fast and empirically close to uniform,
\emph{uniform-like} in \cmsgen{}'s terminology~\cite{golia_cmsgen_2021}, and validates them
with testers such as \barbarik{}~\cite{chakraborty_barbarik_2019}.

\subsection{\bcp{} and Its Data Structures}
\label{sec:related:bcp}

CDCL solvers implement \bcp{} with the 2WL scheme~\cite{moskewicz_chaff_2001}
(Section~\ref{sec:prelim}). 2WL was a breakthrough in \bcp{}
implementation, yet its cost remains a bottleneck~\cite{zhang_cache_2004,iser_unit_2021},
motivating further refinements: cache-conscious layouts of the clause
database~\cite{chu_cache_2009}, watches placed on the literals that stay
satisfied longest~\cite{iser_unit_2021}, and watch lists built as linked structures, which a recent re-evaluation found not to pay
off~\cite{coutelier_link_2024}.
All of them keep the 2WL's core update: falsifying a watched literal moves the watch to
another literal of the clause, a data-dependent update that a CPU handles well and a GPU does
not.

The counter-and-key bookkeeping \gpap{} builds on (Section~\ref{sec:prelim}) goes back to
\nanosat{}~\cite{biere_evolution_2004} and local search~\cite{balint_improving_2014}, and we
revisited it for local search in \noblecount{}~\cite{noblecount_fmcad_2026}, noting that its
uniform, branch-free update suits parallel, GPU-based clause-state maintenance. \gpap{} realizes
and improves upon that suggestion, replacing the XOR with a sum, which, unlike the XOR, lets the counter and the
key update together in a single atomic.

\subsection{Propagation and SAT on GPU}
\label{sec:related:gpu}

Most work that brings propagation to the GPU parallelizes a \emph{single} propagation call.
In SAT, Dal Pal\`u et al.\ and Fujii and Fujimoto spread one \bcp{} call across the clauses and
measure against their own CPU propagators, which scan the clauses for unit and conflicting ones
rather than use watched literals, reporting up to
$15\times$~\cite{dalpalu_cudasat_2015} and $6.7\times$~\cite{fujii_gpu_2012}, respectively. The
same shape appears beyond SAT: domain propagation over the linear constraints of a mixed-integer
program runs on the GPU~\cite{sofranac_domprop_2022}, and so does the propagator of one
global constraint, such as AllDifferent, inside a CPU constraint
solver~\cite{tardivo_gpuprop_2023}. In SAT, a branch-free probing algorithm that propagates
many probes at once, bit-parallel, was designed for GPU portability but remained a CPU
study~\cite{aistleitner_bitparallel_2013}.

Other SAT efforts leave propagation alone. Meyer et al.\ parallelize search over random
3-SAT and concede that sequential solvers are faster by orders of magnitude~\cite{meyer_3sat_2010}. The rest
accelerate components around a CDCL search whose \bcp{} stays on the CPU: look-ahead
scores~\cite{collevati_gpu_2022}, the clauses portfolio solvers exchange~\cite{prevot_gpushare_2021},
and simplification and inprocessing~\cite{osama_parallel_2019,osama_sat_2021}.
\turbosat{}~\cite{dai_turbosat_2025} and \fastfouriersat{}~\cite{cen_fastfouriersat_2024} drop
propagation on the GPU and relax the formula to find one satisfying assignment, the same goal
as CDCL; \turbosat{} completes its best partial assignments with CDCL on the CPU.
None of the SAT efforts above reports a GPU \bcp{} outperforming the 2WL scheme on a CPU.

Closest to \gpap{} is \turbo{}~\cite{talbot_gpucp_2022,talbot_gpucp_2026}, a complete solver
for discrete constraint satisfaction and optimization that runs propagation and
backtracking search on the GPU. Like \gpap{}, it keeps many independent subproblems in flight,
but one per thread would take too much memory, so each gets a whole thread block
(eight warps on one SM), whose threads propagate all of its constraints in parallel until no
domain changes~\cite{talbot_gpucp_2026}; a GPU thus runs only a few hundred at once.

\section{The \gpap{} Engine} 
\label{sec:engine} 

\begin{figure}[t]
\centering
\begin{tikzpicture}[font=\footnotesize]
  \node[draw,rounded corners,fill=black!5,text width=2.5cm,align=center,minimum height=2.0cm]
    (f) at (1.4,1.1) {\textbf{The formula}\\[2pt] occurrence lists\\ [2pt]
                       \emph{one shared,}\\ \emph{read-only copy}};
  \draw[rounded corners,fill=black!4,draw=black!25] (3.8,-0.35) rectangle (11.5,1.8);
  \draw[rounded corners,fill=black!6,draw=black!35] (3.55,-0.15) rectangle (11.25,2.0);
  \draw[rounded corners,fill=black!8,draw=black!70,thick] (3.3,0.05) rectangle (11.0,2.2);
  \node[font=\small\bfseries] at (7.15,1.95) {job $i$: private bookkeeping};
  \node[draw,rounded corners,fill=white,align=center,text width=2.7cm,minimum height=1.2cm]
    (cs) at (4.9,0.9) {\textbf{$\langle$counters, keys$\rangle$}\\[2pt]one pair per clause};
  \node[draw,rounded corners,fill=white,align=center,text width=1.9cm,minimum height=1.2cm]
    (as) at (7.7,0.9) {\textbf{assignment}\\[2pt]one bit per literal};
  \node[draw,rounded corners,fill=white,align=center,text width=1.8cm,minimum height=1.2cm]
    (tr) at (9.85,0.9) {\textbf{trail}\\[2pt]pending literals};
  \draw[-{Stealth[length=4pt]}] (f.east) -- (3.3,1.1);
  \node[font=\small] at (7.4,-0.75) {$\times\,\nrows$ jobs per launch};
\end{tikzpicture}
\caption{\gpap{}'s architecture: one shared, read-only formula, and \nrows{} jobs propagating
         against it in one launch, without communicating. Each job holds a private
         copy of a counter-and-key pair per clause, an assignment, and a trail.}
\label{fig:rows}
\vspace{-8pt}
\end{figure}
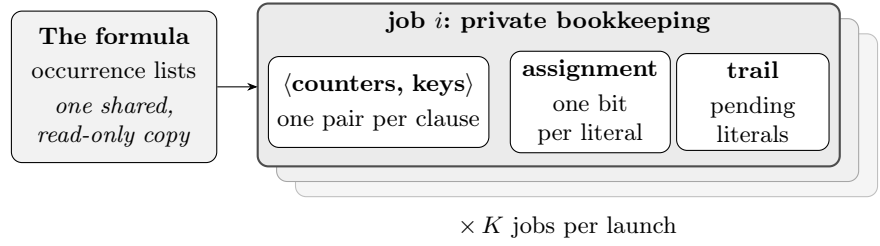

The core technical contribution of this work is \gpap{}, a GPU engine for propagating many independent partial assignments concurrently against a shared CNF formula. \gpap{} implements one primitive: given a fixed CNF formula $F$ and $\nrows$ partial assignments $\litset_1,\dots,\litset_{\nrows}$, each possibly empty, it propagates each $\litset_i$ to its \bcp{} fixpoint independently of the others (Fig.~\ref{fig:rows}) and reports, per assignment, whether a conflict was derived and, otherwise, which literals were implied. On request, it also decides: it picks free variables, assigns them, and propagates.

\subsection{Jobs and the Shared Formula} \label{sec:engine:structure} 

Each partial assignment is a \emph{job}. A job owns a private copy of everything propagation mutates: its assignment, its \emph{trail} (the queue of assigned but not yet processed literals), and the counter and key of every clause (Section~\ref{sec:prelim}). Since different jobs assign different literals, none of this state can be shared. Running many jobs concurrently therefore makes the design of the per-clause bookkeeping just as important as its size. What all jobs do share is the read-only state: the negated occurrence lists $\occ$, stored back-to-back in one array with per-literal offsets, and the initial counter and key of every clause, constructed once. 

\subsection{How Many Jobs Fit} 
\label{sec:engine:unit} 

The amount of private state carried by each job determines how many jobs can run concurrently. We now compute $\nrows$, the number of jobs per kernel launch. A job's bookkeeping occupies device memory for the duration of the launch, so $\nrows$ is bounded by the available memory divided by the bookkeeping required by a single job. Each job occupies one warp, and a GPU keeps only a few thousand warps resident at once, scheduling others as resident warps finish; a large $\nrows$ therefore helps keep the device saturated (Section~\ref{sec:prelim:gpu}). Compactness calls for as few bytes per clause as possible, so we pack a clause's counter and key into a single 32-bit word: the low $\shift$ bits store the counter and the remaining $32-\shift$ bits the key. The key is read only when a single literal survives, at which point it equals that literal's index. The high field therefore needs room for one literal index, and the low field for the largest clause counter. The split $\shift$ is chosen per formula. Literal indices are below $2\nvars$, so when $2\nvars \le 2^{16}-1$ every index fits in 16 bits and we use $\shift = 16$, leaving a 16-bit counter. Larger formulas use $\shift = 8$: the key field then has 24 bits, enough for indices below $2^{24}$, while the counter has 8 bits, capping clause length at $255$. A formula that fits neither split is rejected. Circuits fit easily, since Tseitin encodings produce only short clauses. 

A job's bookkeeping then consists of one such word per clause of three or more literals (4 bytes; binary clauses need none, since falsifying one literal immediately implies the other), an assignment bitmap ($2\nvars$ bits), and a trail --- a queue requiring at most one 4-byte slot per variable, since a variable is assigned at most once before the job either drains the trail or reaches a conflict. Thus 

\begin{equation} \nrows \;\lesssim\; \frac{\text{available device memory}}{4\nclauses \;+\; 2\nvars/8 \;+\; 4\nvars}, \label{eq:budget} \end{equation} where $\nclauses$ counts only clauses of three or more literals. Per the compactness property, bytes per clause directly determine how many jobs fit, which is why we keep the per-clause state to 4 bytes by design. Substituting $\nclauses = 32{,}000$ into Eq.~(\ref{eq:budget}) yields a per-clause term of $4\nclauses = 128$\,KB per job, so a 24\,GB GPU can accommodate on the order of $10^{5}$ partial assignments per launch. The value of $\nrows$ is computed from Eq.~(\ref{eq:budget}) after reserving memory for the shared formula and the device buffers used to return results to the host. Consequently, larger-memory devices support proportionally larger values of $\nrows$, up to a fixed cap.

\subsection{The Sum-Packed Word} 
\label{sec:engine:word} 

Running each job as one warp 
exposes parallelism across jobs. To exploit the GPU more fully, \gpap{} also parallelizes propagation within a job: it spreads the work across the $\wlanes = 32$ lanes of the job's warp, a scheme we call \wave{} (Algorithm~\ref{alg:propagate}). The lanes pop up to $\wlanes$ trail literals at once, concatenate their negated occurrence lists into a single list of clauses to update, and process it $\wlanes$ clauses at a time, one per lane. This is especially important for large formulas, where Eq.~(\ref{eq:budget}) permits fewer concurrent jobs. 

Two lanes may then falsify different literals of the same clause in the same step, and the clause's counter and key must move together, in a single atomic instruction. With the XOR key of Section~\ref{sec:prelim}, they cannot: decrementing the counter is arithmetic, whereas updating the key is bitwise, and no hardware atomic performs both operations at once. Performing them as two atomics costs extra memory traffic and, more importantly, loses the property that a single returned value describes the clause's entire state transition. 

We therefore replace the XOR by a sum, an alternative already mentioned by Balint et al.~\cite{balint_improving_2014}. A sum has the same two properties the key requires: falsifying a literal removes its index, and when a single literal survives, the sum equals that literal's index. Both updates are now subtractions, so they combine into a single atomic operation. \gpap{} therefore keeps as its key $\litsum$, the sum of the indices of the not-yet-falsified literals modulo $2^{32-\shift}$, and packs it with the counter $\satcount$ into one \emph{\sumword{}} $\pkt = (\litsum \ll \shift) + \satcount$. 

Since $\litsum$ is maintained modulo $2^{32-\shift}$, overflow simply wraps around. This is harmless: $\litsum$ is read only when a single literal survives, at which point it equals that literal's index, which fits the field by construction.

Falsifying $\neg\ell$ decrements $\satcount$ and subtracts $\neg\ell$'s index from $\litsum$. Both happen in one arithmetic operation on the word: 

\begin{equation}
  \pkt \;\mathrel{-}=\; 1 + \bigl(\neg\ell \ll \shift\bigr).
  \label{eq:update}
\end{equation}

The $-1$ never borrows from the high field: $\satcount$ counts surviving literals, and each is falsified at most once, so $\satcount$ never becomes negative. The high-field subtraction wraps modulo $2^{32-\shift}$, preserving $\litsum$ in that domain. Thus, when $\satcount$ reaches $1$, \emph{the high field is the forced literal}.

Equation~(\ref{eq:update}) subtracts a constant, which the hardware implements as an atomic add of the negated value, and GPU atomics return the previous word. That returned word, $\mathit{old}$, determines every case: $\mathit{old}.\satcount = 1$ means this update falsified the last survivor, producing a conflict; $\mathit{old}.\satcount = 2$ means the clause became unit, and the forced literal is $(\mathit{old}.\litsum - \neg\ell) \bmod 2^{32-\shift}$, since the returned field still holds the sum of the two surviving literals and we know which one was removed; larger values mean the clause remains unresolved. 

When two lanes falsify two literals of the same clause simultaneously, the memory system serializes the atomics and each lane receives the word as it existed immediately before its own update. Every transition, to unit or to conflict, is therefore observed by exactly one lane. This atomics-over-locks property ensures correctness and lets propagation spread across all lanes of a warp.

\subsection{The Assignment: One Bit per Literal} \label{sec:engine:bitmap} 

A job's assignment is a bitmap of $\lceil 2\nvars/32 \rceil$ words in which bit $\ell$ is set iff literal $\ell$ is true. The two polarities of a variable are the adjacent indices $2(v-1)$ and $2(v-1)+1$, so they always reside in the same word. Assigning a forced literal $u$ is therefore one \textsc{AtomicOr}, whose returned word answers the two questions that matter at once: bit $u$ already set means that $u$ is already assigned, and bit $u \oplus 1$ already set means that the opposite polarity is already assigned, i.e., a conflict. Two lanes forcing $u$ and $\neg u$ in the same step are simply ordered by the hardware, and exactly one of the two returned words shows the other's bit, so exactly one lane reports the conflict. The bitmap is also eight times smaller than a byte-per-literal representation, helping keep assignments in cache.

\begin{algorithm}[t]
\caption{\wave{}: propagate one job to fixpoint across $\wlanes$ lanes}
\label{alg:propagate}
\begin{algorithmic}[1]
\While{the trail is non-empty}
  \State pop a literal $\ell$ from this job's trail \Comment{up to $\wlanes$ literals at once, one per lane}
  \ForAll{clauses $c \in \occ[\ell]$} \Comment{$\wlanes$ clause updates per step}
    \State $\mathit{old} \gets \textsc{AtomicAdd}\bigl(\pkt[c],\, -(1 + (\neg\ell \ll \shift))\bigr)$
    \If{$\mathit{old}.\satcount = 1$} \State \textbf{report} conflict; \textbf{stop this job}
    \ElsIf{$\mathit{old}.\satcount = 2$}
      \State $u \gets (\mathit{old}.\litsum - \neg\ell) \bmod 2^{32-\shift}$
             \Comment{the forced literal}
      \State $\mathit{prev} \gets \textsc{AtomicOr}\bigl(\mathit{bitmap}[\lfloor u/32 \rfloor],\, 1 \ll (u \bmod 32)\bigr)$
      \If{bit $u \oplus 1$ of $\mathit{prev}$ is set} \State \textbf{report} conflict; \textbf{stop this job}
      \ElsIf{bit $u$ of $\mathit{prev}$ is clear} \State push $u$ on the trail
      \EndIf
    \EndIf
  \EndFor
\EndWhile
\end{algorithmic}
\end{algorithm}

\subsection{Two Kernels, One Propagation Core} \label{sec:engine:kernels} The engine exposes two operations, one kernel each. \emph{Propagate to fixpoint} asserts $\nrows$ partial assignments and runs \bcp{} to fixpoint or conflict in every job. \emph{Run a trajectory} executes a whole randomized decide-and-propagate trajectory, from the first decision to a conflict or a full assignment, within a single kernel launch: each job repeatedly takes the next $\ndec$ free variables in a given order, assigns each a polarity drawn from a stateless hash of (seed, job, variable), and propagates. 

Both operations share the propagation core of Algorithm~\ref{alg:propagate}. For formulas with fewer than roughly $2{,}000$ clauses, a job's propagation cannot fully occupy $\wlanes$ lanes, so the engine instead runs it serially within its warp, using ordinary loads and stores rather than atomics. Both variants operate on the same bookkeeping and, since \bcp{} is confluent, reach the same fixpoint.

\subsection{Using the Engine}
\label{sec:engine:api}

The primitive is not specific to the sampler, so we release it as a
library: a C core with thin C++ and Python interfaces over the same implementation. The
kernels are written in \triton~\cite{tillet_triton_2019} and compiled ahead of time into the
library.
Checking many partial assignments at
once is two lines:

\begin{center}
\begin{minipage}{0.86\linewidth}
\begin{lstlisting}[basicstyle=\ttfamily\footnotesize,frame=single,language=C++,
                   keywordstyle=\bfseries,columns=fullflexible]
gpap::AssumptionChecker checker("formula.cnf");
// any number of assignments (DIMACS literals), K per launch
auto r = checker.check({{1,-2,3}, {-1,2,-3}, ...});
\end{lstlisting}
\end{minipage}
\end{center}

\noindent
The trajectory kernel is exposed analogously. Using \gpap{} therefore amounts to an API call and requires no GPU programming.

\section{\gpaps{}: GPU-based SAT Sampling}
\label{sec:sampling}

We now build a sampler on top of \gpap{}, exploiting its ability to execute many independent decide-and-propagate trajectories concurrently on a shared formula. Our sampler takes a \cnf{} formula together with a support $\support$, and draws projected solutions by running many randomized decide-and-propagate trajectories.

\subsection{The Sampling Algorithm} \label{sec:sampling:algo} 

\gpaps{} samples by exploring many randomized decide-and-propagate trajectories in parallel and retaining those that reach models. A trajectory starts from the empty assignment and repeats one decision round: decide the next $\ndec$ free variables, giving each a random polarity, then propagate to \bcp{} fixpoint (Section~\ref{sec:engine:kernels}). Decisions take the support variables first; if the fully assigned support does not determine the rest through propagation, as it can fail to for a general \cnf{}, the trajectory keeps deciding among the remaining variables. For circuits the support is the primary inputs, which determine every other variable, so no decision is needed beyond the support. A trajectory can still conflict: it decides only $\ndec$ inputs per round, and propagating them can contradict values that the pinned target and earlier rounds already forced. A trajectory that conflicts is discarded; one that assigns every variable is a model, and is read out, restricted to $\support$, as one sample. The engine runs $\nrows$ such trajectories per launch against the shared formula, so each launch returns up to $\nrows$ samples.

\subsection{The Sampling Pipeline}
\label{sec:sampling:pipeline}

\begin{figure}[t]
\centering
\begin{tikzpicture}[font=\footnotesize,
    box/.style={draw,rounded corners,fill=black!6,align=center,inner sep=2.5pt,
                text width=2.0cm,minimum height=1.35cm},
    eng/.style={draw,rounded corners,fill=black!14,align=center,inner sep=4pt},
    ar/.style={-{Stealth[length=4pt]}},
    use/.style={{Stealth[length=4pt]}-{Stealth[length=4pt]},densely dashed}]
  \node[box] (in)   at (0,0)     {\textbf{input}\\[2pt]\cnf{} formula,\\ support $\support$};
  \node[box] (mine) at (2.46,0)  {\textbf{mine}\\[2pt]entailed\\ clauses};
  \node[box] (simp) at (4.92,0)  {\textbf{simplify}\\[2pt]units,\\ equivalences};
  \node[box] (samp) at (7.38,0)  {\textbf{sample}\\[2pt]configuration\\ chosen online};
  \node[box] (out)  at (9.84,0)  {\textbf{project}\\[2pt]bit matrix\\ over $\support$};
  \node[eng,text width=7.0cm] (eng) at (4.92,-2.1) {\gpap{} engine (Section~\ref{sec:engine}):
        $\nrows$ jobs per launch};
  \draw[ar] (in) -- (mine);
  \draw[ar] (mine) -- (simp);
  \draw[ar] (simp) -- (samp);
  \draw[ar] (samp) -- (out);
  \draw[use] (mine.south) -- node[midway,fill=white,inner sep=1.5pt,font=\scriptsize,
        align=center] {$\nrows$ literal sets\\per launch} (mine.south |- eng.north);
  \draw[use] (samp.south) -- node[midway,fill=white,inner sep=1.5pt,font=\scriptsize,
        align=center] {$\nrows$ trajectories\\per launch} (samp.south |- eng.north);
\end{tikzpicture}
\caption{\gpaps{}'s sampling pipeline; each box is a stage (bold) and what it handles. Mining and configuration choice both ride the engine's batch parallelism: a
         mining launch probes $\nrows$ candidate literal sets at once, and each
         configuration is measured on a launch of $\nrows$ trajectories. Simplification
         preserves the projected solution set exactly, and projection hands the host a
         bit matrix over $\support$.}
\label{fig:pipeline}
\vspace{-8pt}
\end{figure}
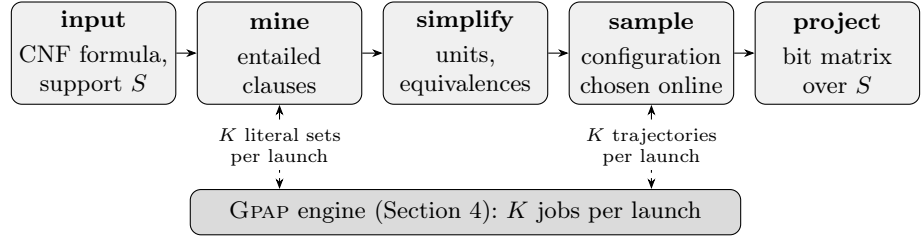

\gpaps{} uses \gpap{} not only for sampling but also for formula preprocessing and online configuration.
Figure~\ref{fig:pipeline} shows the pipeline. \gpaps{} first mines the input \cnf{} for clauses
it entails, then simplifies the augmented formula, then samples with the loop of
Section~\ref{sec:sampling:algo}, its configuration chosen online, and
finally projects the samples onto $\support$ on the device. Mining and configuration choice
are themselves applications of \gpap{}: a mining
launch probes $\nrows$ literal sets at once, and each configuration is measured on a
full launch of $\nrows$ trajectories.

\paragraph{Mining.} Before sampling, \gpaps{} strengthens the formula with cheaply discoverable consequences that can increase propagation and reduce the search space. If assuming a set of literals and propagating leads to a
conflict, the formula entails the clause made of their negations. For a single literal
this is failed-literal probing~\cite{lynce_probing_2003}, and the entailed unit is a
\emph{backbone} literal, one that holds in every model~\cite{janota_backbones_2015}.
\gpaps{} runs such probes over the support literals under a small fixed time
budget: first every single literal, repeated with the units found asserted until no new one
appears, then every pair, then triples as the budget allows. Besides some of the backbone
literals, this mines binary and ternary clauses, which the simplification stage below uses.
The probes are independent of one another, so where a CPU must run
them as a loop of propagation passes, the engine runs $\nrows$ of them in parallel as one
launch, one probe per job. The mined clauses are added to the formula and kept only if a
short warm-up run completes more trajectories with them than without.

\paragraph{Simplification.} 
The mined information is then used to reduce the formula while preserving the projected solution set. The formula (possibly augmented with the mined clauses) is
simplified: its units are asserted, and equivalent-literal substitution collapses variables
that were proven equal (or opposite) by pairs of binary clauses. The
substitutions are recorded so every sample can be reconstructed at readout. Both steps preserve the projected solution set
exactly; preprocessing such as variable elimination can change it, and \cmsgen{} likewise turns
variable elimination off~\cite{golia_cmsgen_2021}.

\paragraph{Choosing a Configuration.} 
Different instances favor different sampling configurations, so \gpaps{} selects among them online. A \emph{configuration} of the sampling loop is the
decision order together with $\ndec$ (the number of variables decided per round), and none wins everywhere: for circuits
\gpaps{} runs random order with $\ndec = 8$ and with $\ndec = 1$. Choosing between
them online is a multi-armed bandit
problem~\cite{slivkins_bandits_2019,lattimore_bandits_2020}, a formulation also used for
branching in CDCL solvers~\cite{cherif_kissatmab_2021} and
local-search MaxSAT~\cite{zheng_bandmaxsat_2022,zheng_bandhs_2022}: each configuration is an
arm, and one `pull' runs it for a full launch, so
one pull already measures $\nrows$ independent runs of the loop. After one pull of each arm,
\gpaps{} favors the arm with the highest throughput.

\paragraph{Projection.} The run keeps its samples on the device. When they are written out, all that crosses to the host is a bit matrix over $\support$, one row per sample, projected on the GPU, so the full $\nvars$-variable assignment never leaves the device.

\subsection{Using \gpaps{}}
\label{sec:sampling:usage}

\gpaps{} is available as the \texttt{sample} command of \texttt{gpap} and as a Python API. It accepts a DIMACS \cnf{}, projected onto its \texttt{c ind} variables (or all variables if absent), or a \texttt{.bench} or AIGER circuit, which it unrolls, pins, and encodes:

\begin{center}
\begin{minipage}{0.86\linewidth}
\begin{lstlisting}[basicstyle=\ttfamily\footnotesize,frame=single,columns=fullflexible]
gpap sample formula.cnf -n 50000 --out samples.txt
gpap sample b10.bench --cycles 25 -n 50000
\end{lstlisting}
\end{minipage}
\end{center}

\noindent
A run stops once it has the requested number of samples, or at its time limit (60\,s by
default), and writes one row per sample over the support.

\subsection{Sample Uniformity} \label{sec:sampling:exactness} Because \gpaps{} provides no formal uniformity guarantee, we evaluate it as \cmsgen{} was evaluated~\cite{golia_cmsgen_2021}, with \barbarik{}~\cite{chakraborty_barbarik_2019} and the same parameters. 

Given a uniform \emph{reference} sampler, here \spur{}~\cite{achlioptas_spur_2018}, \barbarik{} decides, with confidence $1-\delta$, whether the \emph{sampler under test} is $\varepsilon$-close to uniform (accept) or $\eta$-far from it (reject). It repeatedly draws one solution from each sampler, builds a formula whose projected solutions are exactly those two samples, and rejects if the sampler under test draws them in a ratio far from even. 

These parameters, $\varepsilon = 0.3$, $\eta = 1.8$, and $\delta = 0.1$, require $1{,}836$ samples from the sampler under test per instance; we run these tests on an NVIDIA RTX 4080 Laptop GPU. The instances come from the circuit families of Section~\ref{sec:eval} but form a separate, smaller set: 12 sequential circuits unrolled for 5, 10, and 25 cycles and 10 ISCAS-85 circuits (all but \texttt{c17}), each pinned to both a random and a rare target. Of the 92 resulting instances, we keep the 65 on which \barbarik{} terminates and accepts \spur{} when \spur{} itself is placed under test. 

\gpaps{} is accepted on all 65 instances, as is \cmsgen{} under the same test.

\section{Experimental Evaluation} 

\label{sec:eval} 

We evaluate \gpaps{} along the two dimensions that motivate this work: throughput and the ability to sample rare targets. We compare \gpaps{} against two kinds of baselines: \cmsgen{}, the CPU-based SAT sampler, and the GPU gradient samplers, both on the latter's benchmark families and at rare targets.

\subsection{Benchmarks and Protocol}
\label{sec:protocol}

\runin{Setup.} All tools run on one machine: \gpaps{} and the GPU gradient samplers,
\demotic{} and \recurrent{}, on one NVIDIA L4 GPU (24\,GB), and \cmsgen{} on one core of an
Intel Xeon at $2.20\,$GHz. \gpaps{} and \cmsgen{} get a $60\,$s sampling budget per instance,
and a gradient sampler gets $1{,}200\,$s to build and initialize its model, which can take many
minutes over a deep unroll. Of the gradient samplers, we measure the two built for circuits,
and as our CPU baseline we take \cmsgen{}, the best-performing CPU sampler in both of their
evaluations~\cite{ardakani_demotic_2025,ardakani_recurrent_2025}.

\runin{Benchmarks.} We take two sets from the gradient samplers' own papers:
\emph{ISCAS-85}, the 11 combinational ISCAS-85 circuits~\cite{brglez_iscas85_1985} that
\demotic{} evaluates, and \emph{Sequential}, 30 circuits from the families \recurrent{}
evaluates, all 21 of ITC'99~\cite{corno_itc99_2000} and 9 of ISCAS-89~\cite{brglez_iscas89_1989},
each unrolled from reset for 25 and for 50 cycles. Each circuit is pinned to two targets. The
\emph{\orig{} target} fixes only the one to five output bits that the set's gradient sampler
fixes in its own released experiments, and random simulation hits it often. \emph{Our rare
target} fixes every final-cycle output to the rare target of Section~\ref{sec:prelim}, observed
over $262{,}144$ random input sequences; random simulation hits it with a median probability of
about $10^{-5}$, though on circuits with few distinct output patterns even the rarest is hit
much more often.
This gives $142$ (circuit, target, depth) instances. \abctool{}~\cite{brayton_abc_2010}, which
our encoder runs, crashes on one of them; the other $141$ encode up to $1.1\e{6}$ clauses, and even the largest leaves room for $\nrows = 5{,}161$ trajectories per launch.

\runin{Runs.} Every sampler receives the same circuit, simplified with
\abctool{},\footnote{\abctool{} runs \texttt{dc2; dc2; fraig} before unrolling; \gpaps{}'s
encoder repeats it and adds a \texttt{dc2} after.} and each gradient sampler runs only on its own
set, once, with its own script's default settings and the largest batch it can run to
completion; a run that times out or does not initialize counts as a failure. \recurrent{} counts
a sequence as a solution if the target holds at any depth up to a bound, so we give every tool
the same depth: \recurrent{} the bound $\depth$, counting its solutions over all depths as its
paper does~\cite{ardakani_recurrent_2025}, and \gpaps{} and \cmsgen{} exactly $\depth$ cycles.
These two sample the same Tseitin-encoded~\cite{tseitin_complexity_1983} \cnf{} with a
$50{,}000$-sample request, \gpaps{} with its default settings and the median of three seeds;
since it samples whole launches of $\nrows$ trajectories, it often delivers more, and we count
every sample. No sampler's clock includes collecting and writing out samples: we patch \cmsgen{}
to generate without writing, and deduct the time \recurrent{} spends copying its samples to the
host inside its timed loop.

\runin{Metrics.} We report samples per second on two clocks: \emph{steady state} (ss) times
the sampling run alone, and \emph{end to end} (e2e) times \gpaps{}'s whole run, from parsing the
formula to the last sample. Against \cmsgen{} we report both and count every sample; \cmsgen{}
starts sampling as soon as it has read the formula, so its two rates are practically equal, and
Table~\ref{tab:cmsgen} gives it a single column. Against the gradient samplers we report steady
state only, since their published rates exclude model build and initialization, and count only
unique samples, for both tools, as their papers do~\cite{ardakani_demotic_2025,ardakani_recurrent_2025}; duplicates are removed after the timed run.

\runin{Filters.} We exclude from every geometric mean the instances that do not measure
sampling, each detected mechanically: \emph{vacuous} targets (the target holds on every input),
\emph{free-support} instances (after propagation every assignment to the support is a
solution), \emph{tiny} spaces (fewer than $1{,}000$ solutions, which both tools enumerate), and
instances proved \emph{unsatisfiable} at their depth. We also exclude the instances where
neither sampler emits and the one \abctool{} crashes on.

\subsection{Comparison against \cmsgen{}}
\label{sec:eval:cmsgen}

\begin{table}[!t]
\centering
\caption{\gpaps{} vs.\ \cmsgen{}: samples per second, geometric means over the $n$ instances
         where both emit, for \gpaps{} in steady state (ss) and end to end (e2e). ISCAS-85 is
         \demotic{}'s benchmark set (11 instances per row), Sequential \recurrent{}'s (30). C$=$0: instances where only \gpaps{} emits; Wins: instances where \gpaps{} is faster, out of $n$ plus C$=$0.}
\label{tab:cmsgen}
\footnotesize
\setlength{\tabcolsep}{4pt}
\begin{tabular}{llrrrrrrr}
\toprule
& & & & \multicolumn{3}{c}{samples/s} & & \\
\cmidrule(lr){5-7}
Set & Target & $\depth$ & $n$ & \cmsgen{} & \gpaps{} ss & \gpaps{} e2e & Wins & C$=$0 \\
\midrule
ISCAS-85   & \orig{} & 1  & 10 & $1.78\e{3}$ & $2.37\e{6}$ & $8.26\e{5}$ & $10/10$ & 0 \\
ISCAS-85   & rare    & 1  & 7  & $1.10\e{4}$ & $2.65\e{6}$ & $6.77\e{5}$ & $7/7$   & 0 \\
Sequential & \orig{} & 25 & 16 & $323$       & $2.80\e{5}$ & $1.11\e{5}$ & $23/23$ & 7 \\
Sequential & rare    & 25 & 18 & $811$       & $6.76\e{4}$ & $2.51\e{4}$ & $19/22$ & 4 \\
Sequential & \orig{} & 50 & 15 & $194$       & $3.04\e{5}$ & $9.56\e{4}$ & $25/25$ & 10 \\
Sequential & rare    & 50 & 16 & $369$       & $3.16\e{4}$ & $1.35\e{4}$ & $21/23$ & 7 \\
\bottomrule
\end{tabular}
\vspace{-6pt}
\end{table}

\begin{figure}[!t]
\centering
\includegraphics[width=0.8\linewidth]{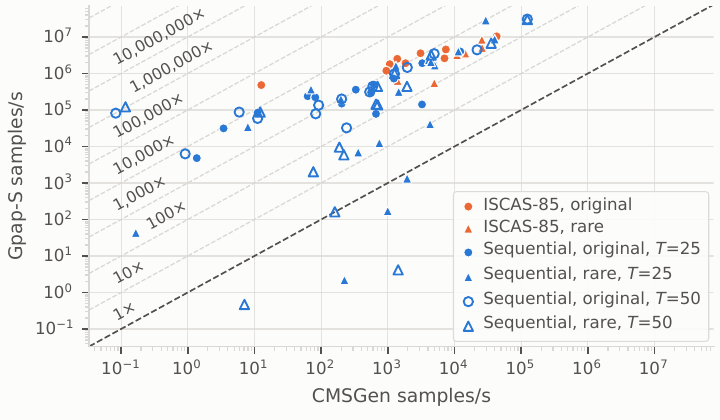}
\caption{Samples/s, \gpaps{} (steady state) against \cmsgen{}, one dot per instance of Table~\ref{tab:cmsgen} where both emit. Circles denote \orig{} targets, triangles rare targets, and hollow markers $\depth=50$; diagonals are constant-ratio lines. Missing dots are C$=$0, filtered, or unsampled instances.}
\label{fig:sps}
\vspace{-8pt}
\end{figure}

Table~\ref{tab:cmsgen} shows the comparison. Each row is one benchmark set under one target at
one depth, and gives each sampler's rate as the geometric mean over the row's $n$ instances
on which both emit.
\gpaps{} is faster on every row, by
$83.4$--$1{,}571\times$ in steady state and $31.0$--$494\times$ end to end.
Of the 110 unfiltered instances where
either emits, \gpaps{} delivers all $50{,}000$ requested samples within its budget on 93,
\cmsgen{} on 44. Figure~\ref{fig:sps} plots every instance where both emit.

The target decides the size of the speedup. At the \orig{} targets, which random inputs hit often, the speedup
is largest, $1{,}335\times$ on ISCAS-85, and on the Sequential set it grows with depth, from $866\times$ at $\depth = 25$
to $1{,}571\times$ at $\depth = 50$, as \cmsgen{}'s work per sample grows faster with depth than
\gpaps{}'s (Section~\ref{sec:eval:throughput}). At these
targets \cmsgen{} also emits nothing at all within its budget on 7 ($\depth = 25$) and 10
($\depth = 50$) sequential instances that \gpaps{} samples; they count as wins but, having no
ratio, stay out of the geometric mean. At the rare target the speedup drops to $240\times$
(ISCAS-85), $83.4\times$ and $85.7\times$ (Sequential): solutions are scarce, and most of
\gpaps{}'s trajectories conflict and are discarded, hence each emitted sample costs more propagation. Even so, \gpaps{} remains at least
$83.4\times$ faster in geometric mean. It is slower on five sequential rare-target instances, where at most $0.41\%$ of its
trajectories complete and \cmsgen{}'s learned clauses pay off; on three ($\depth = 25$) and
four ($\depth = 50$) of the largest sequential circuits, neither sampler emits at the rare target.

End to end, the speedup is a quarter to a half of the steady-state speedup, because \gpaps{}'s wall time
then includes its preparation (parsing, mining, simplification, and warm-up), a median $33$--$67\%$ of it on
these rows. Preparation is a fixed cost, so a larger
sample request moves the end-to-end speedup toward the steady-state one. Neither clock includes copying the samples to the host and
writing them out, which we omit for every tool (Section~\ref{sec:protocol}).

\subsection{Comparison against the GPU Gradient Samplers}
\label{sec:eval:rivals}

\begin{table}[!t]
\centering
\caption{\gpaps{} vs.\ the GPU gradient samplers, each on its own set: \demotic{} on
         ISCAS-85 and \recurrent{} on Sequential, with every tool at the same depth $\depth$
         (Section~\ref{sec:protocol}). Unique samples per second in steady state, geometric means
         over the $n$ instances where both emit. C$=$0: instances where only \gpaps{} emits; Wins: out of
         $n$ plus C$=$0; Ties: instances whose outcome flips across \gpaps{}'s three seeds, not
         counted as wins; Fail: baseline runs with no measurement inside the budget.}
\label{tab:rivals}
\footnotesize
\setlength{\tabcolsep}{4pt}
\begin{tabular}{llrrrrrrrr}
\toprule
& & & & \multicolumn{2}{c}{unique samples/s} & & & & \\
\cmidrule(lr){5-6}
Baseline & Target & $\depth$ & $n$ & Baseline & \gpaps{} & Wins & Ties & C$=$0 & Fail \\
\midrule
\multirow{2}{*}{\demotic{}}
  & \orig{} & 1  & 10 & $1.05\e{6}$ & $2.37\e{6}$ & $8/10$  & 1 & 0 & 0 \\
  & rare    & 1  & 2  & $2.69$      & $2.16\e{6}$ & $7/7$   & 0 & 5 & 0 \\
\midrule
\multirow{4}{*}{\recurrent{}}
  & \orig{} & 25 & 20 & $6.96\e{3}$ & $1.30\e{5}$ & $16/20$ & 0 & 0 & 4 \\
  & rare    & 25 & 15 & $496$       & $7.28\e{4}$ & $18/19$ & 0 & 4 & 3 \\
  & \orig{} & 50 & 16 & $3.53\e{4}$ & $2.84\e{5}$ & $13/16$ & 1 & 0 & 9 \\
  & rare    & 50 & 15 & $692$       & $5.86\e{4}$ & $16/18$ & 1 & 3 & 9 \\
\bottomrule
\end{tabular}
\vspace{-6pt}
\end{table}

Table~\ref{tab:rivals} shows the comparison: steady-state unique samples per second, in one
block per gradient sampler, on its own benchmark set, and one row per target and depth.
\gpaps{}'s rates differ from Table~\ref{tab:cmsgen}'s: they count only unique samples, and each
mean is over the instances where the gradient sampler emits too.

At the \orig{} targets the gradient samplers are fast: on its own instances \demotic{} emits
${1.05\e{6}}$ unique samples per second, against ${2.37\e{6}}$ for \gpaps{}, whose speedup
is modest: $2.2\times$ over \demotic{}, which is faster on one of its ten instances, and $18.7\times$ and $8.0\times$ over
\recurrent{} at $\depth = 25$ and $50$.

\recurrent{}'s throughput grows with its batch, which GPU memory bounds, as it bounds \gpaps{}'s
$\nrows$ (Eq.~(\ref{eq:budget})). On seven large instances at $\depth = 25$ only a small batch
fits on the card, and there \gpaps{}'s speedup is $107\times$, against $7.3\times$ on the
other thirteen. On still larger instances \recurrent{} fails to
initialize within its $1{,}200\,$s budget: four at $\depth = 25$ and nine at $\depth = 50$
(Fail column).

At the rare target the gap widens (Table~\ref{tab:rivals}). A gradient sampler rounds the result
of a continuous search into a Boolean assignment, and when solutions are scarce, few rounded
assignments are solutions. \gpaps{} slows down too, as more of its trajectories conflict, but
far less.
\demotic{} finds no solution on five of its seven rare-target instances, and on the other two it
is over five orders of magnitude slower than \gpaps{}. \recurrent{} still finds solutions, but
\gpaps{}'s speedup over it grows from $18.7\times$ and $8.0\times$ to $147\times$ and
$84.7\times$. 

\subsection{Where the Speed Comes From}
\label{sec:eval:throughput}

Table~\ref{tab:cmsgen}'s speedups range from $83.4\times$ to $1{,}571\times$, and
Table~\ref{tab:prop} explains this spread with two quantities. A sampler's
\emph{propagation rate} is the number of implications it performs per second, an implication
being one literal forced by the unit-clause rule. Its \emph{work per sample} is the number of
implications it spends for each sample it emits, those of discarded trajectories included.
A sampler's rate in samples per second is its propagation rate divided by its work per
sample, so each speedup of Table~\ref{tab:cmsgen} is one ratio divided by another, $\text{speedup} = \text{rate ratio}/\text{work ratio}$, where the \emph{rate ratio} is \gpaps{}'s propagation rate over \cmsgen{}'s, and the
\emph{work ratio} is \gpaps{}'s work per sample over \cmsgen{}'s.

\begin{table}[!t]
\centering
\caption{Breakdown of Table~\ref{tab:cmsgen}'s steady-state speedups, over the same rows and
         instances: propagation rate (\implps{}) and work per sample (implications per
         emitted sample). The work ratio is \gpaps{}'s over \cmsgen{}'s.}
\label{tab:prop}
\footnotesize
\setlength{\tabcolsep}{3.4pt}
\begin{tabular}{llrrrrrrr}
\toprule
& & & & \multicolumn{2}{c}{\implps{}} & \multicolumn{2}{c}{impl./sample} & Work \\
\cmidrule(lr){5-6}\cmidrule(lr){7-8}
Set & Target & $\depth$ & $n$ & \cmsgen{} & \gpaps{} & \cmsgen{} & \gpaps{} & ratio \\
\midrule
ISCAS-85   & \orig{} & 1  & 10 & $1.56\e{6}$ & $1.92\e{9}$ & 876 & 811 & $0.93\times$ \\
ISCAS-85   & rare    & 1  & 7  & $4.85\e{6}$ & $1.59\e{9}$ & 440 & 600 & $1.36\times$ \\
Sequential & \orig{} & 25 & 16 & $4.25\e{6}$ & $2.27\e{9}$ & 13{,}145 & 8{,}106 & $0.62\times$ \\
Sequential & rare    & 25 & 18 & $3.93\e{6}$ & $2.10\e{9}$ & 4{,}849 & 31{,}085 & $6.41\times$ \\
Sequential & \orig{} & 50 & 15 & $4.29\e{6}$ & $2.20\e{9}$ & 22{,}177 & 7{,}233 & $0.33\times$ \\
Sequential & rare    & 50 & 16 & $4.01\e{6}$ & $2.27\e{9}$ & 10{,}869 & 71{,}607 & $6.59\times$ \\
\bottomrule
\end{tabular}
\vspace{-6pt}
\end{table}

\gpaps{}'s propagation rate is very similar across the rows. It sustains $1.59\e{9}$ to $2.27\e{9}$ implications per
second on every row, so the slowest row is within $1.43\times$ of the fastest, and their
geometric mean is $568\times$ \cmsgen{}'s. The rate ratio therefore tracks \cmsgen{}'s
propagation rate. On the Sequential set that rate barely changes either, so the rate ratio
stays between $512\times$ and $565\times$ and the work ratio accounts for the spread; on
ISCAS-85, \cmsgen{} propagates three times faster at the rare target, cutting the rate ratio
from $1{,}237\times$ to $328\times$.

The work ratio swings with the target. At the rare target on the Sequential set, \gpaps{} needs
$6.41\times$ ($\depth = 25$) and $6.59\times$ ($\depth = 50$) \cmsgen{}'s work per sample:
most of its trajectories conflict and are discarded with all their work,
whereas \cmsgen{} learns a clause from each conflict and backtracks, so it does not repeat
the conflict. Those two rows are exactly
Table~\ref{tab:cmsgen}'s smallest speedups, $83.4\times$ and $85.7\times$. At the \orig{} targets the
relation inverts: at $\depth = 50$, \gpaps{} needs only $0.33\times$ \cmsgen{}'s work
per sample, hence Table~\ref{tab:cmsgen}'s largest speedup.

\section{Conclusion}
\label{sec:conclusion}

We presented \gpaps{}, a GPU sampler built on \bcp{}, and \gpap{}, the engine behind it, which
propagates many independent partial assignments at once against one shared formula. Together,
they bring the rare-target robustness of SAT-based sampling to GPU throughput. \gpaps{} discards conflicting trajectories rather than learning from them, which leaves room for improvement where few trajectories complete: for instance, sharing learned clauses across trajectories.

\gpap{} itself is a general \bcp{} engine, and Table~\ref{tab:prop} shows its raw power: on one
GPU, it performs two to three orders of magnitude more implications per second than a CDCL
solver on one CPU core. Since CDCL solvers spend most of their time in \bcp{}, we expect \gpap{}
to find many uses beyond sampling: wherever many assignments
are propagated against one formula, from probing, SAT sweeping, and look-ahead in solvers to
SAT-based tools such as preprocessors, model checkers, and test generators.

\section*{Data Availability Statement}
\gpaps{} and \gpap{} are open source under the MIT license at
\url{https://github.com/yogevshalmon/gpap/tree/v0.3.0}.

\bibliographystyle{splncs04}
\bibliography{paper}

\end{document}